\documentstyle[preprint,aps]{revtex}
\newcommand{\be}{\begin{equation}}
\newcommand{\ee}{\end{equation}}
\newcommand{\ba}{\begin{eqnarray}}
\newcommand{\ea}{\end{eqnarray}}

\begin{document}

\draft

\preprint{October,2000}

\title{Non-vanishing Magnetic Flux through the Slightly-charged \\
       Kerr Black Hole}

\author{Hongsu Kim\footnote{e-mail : hongsu@hepth.hanyang.ac.kr},
Chul Hoon Lee\footnote{e-mail : chlee@hepth.hanyang.ac.kr} and
Hyun Kyu Lee\footnote{e-mail : hklee@hepth.hanyang.ac.kr}}

\address{Department of Physics \\
Hanyang University, Seoul, 133-791, KOREA}


\maketitle

\begin{abstract}
In association with the Blanford-Znajek mechanism for rotational energy extraction
from Kerr black holes, it is of some interest to explore how much of magnetic flux
can actually penetrate the horizon at least in idealized situations. For completely
uncharged Kerr hole case, it has been known for some time that the magnetic flux
gets entirely expelled when the hole is maximally-rotating. In the mean time, it is
known that when the rotating hole is immersed in an originally uniform magnetic field
surrounded by an ionized interstellar medium (plasma), which is a more realistic
situation, the hole accretes certain amount of electric charge. In the present work,
it is demonstrated that as a result of this accretion charge small enough not to
disturb the geometry, the magnetic flux through this slightly charged Kerr hole depends
not only on the hole's angular momentum but on the hole's charge as well such that
it never vanishes for any value of the hole's angular momentum.
\end{abstract}


\narrowtext

\newpage
\begin{center}
{\rm\bf I. Introduction}
\end{center}

Among various models attempting to provide a consistent account for the spectrum
of quasars, radio galaxies or the gamma ray bursters in which 
strong magnetic field is believed to be 
anchored in the central black hole, perhaps the mechanism proposed by Blanford
and Znajek [1] might be the most attractive and natural one to theoretical physicists.
Despite the skepticism generally held by astronomers and majority of astrophysicists,
the Blanford-Znajek mechanism for the extraction of rotational energy from rotating
black holes has remained an issue of great interest in theoretical astrophysics community
due to its concreteness in formulation and plausibility in operational nature. 
Being so, there has been continuous research activities 
to ask and answer questions relevant to the environment set by the Blanford-Znajek
mechanism such as the stationary, axisymmetric magnetosphere in which the rotating
hole is surrounded by a strong magnetic field and plasma. Among such
questions, one of the most interesting thought experiments which can be explored
in an analytic manner at least in idealized situations is the issue of how much
of the magnetic field can actually penetrate the rotating hole's horizon. Or more
precisely, through one half of the surface of the horizon as the total flux across
the whole horizon should be zero. The right answer to this question can indeed be
crucial in order for the Blanford-Znajek mechanism and its generalized versions [2]
to work at all as they all rely on the picture in which magnetic fields (particularly, their
poloidal components) penetrating the hole's horizon and the surrounding accretion
disk, transmit the rotational energy of the hole to distant relativistic particles
and fields. Interestingly but rather to our dismay, it has been found [3,10] that this
magnetic flux intersecting the horizon decreases as the angular momentum of the
hole increases and goes to zero when the hole is maximally rotating. Considering
that larger amount of rotational energy can be extracted from more rapidly rotating
black holes, this result, even if we take into account the idealized setup employed,
is quite adverse to the operational aspect of the Blanford-Znajek mechanism.
Because of this negative implication of the result, there have also been some
other works employing more careful treatments to turn the conclusion around
and hence save the mechanism.
The work performed by Dokuchaev [4] employing the Ernst-Wild solution [5]
to the coupled Einstein-Maxwell equations belongs to this category. Starting from the
exact Ernst-Wild solution, he showed that the magnetic flux through the hole 
becomes independent of the hole's angular momentum when the hole builds up 
equilibrium (Wald) charge. In the present work, we
choose to deal with the issue more directly and demonstrate in a transparent manner
that in a more realistic situation when there are charges around (but small enough
not to disturb the background vacuum geometry), the magnetic flux penetrating the horizon
depends not only on the angular momentum but on the amount of charge accreted on the hole 
as well and can never vanish no matter how fast the hole spins. In order to find the 
electromagnetic field around a rotating black hole and hence eventually the magnetic 
flux through one half of the horizon, we look for the solution to the source-free
Maxwell equations in the background of the stationary, axisymmetric vacuum spacetime,
the Kerr geometry. And perhaps the most concise and elegant way to obtain this solution
which occurs when a stationary, axisymmetric black hole is placed in an asymptotically
uniform magnetic field would be the simple algorithm suggested long ago by Wald [6]
which is based upon the realization [7] that the Killing vectors owned by a given vacuum
spacetime generate solutions of source-free Maxwell equations in the background of
that spacetime. Here, although ``vacuum'' situation represents the case when both
the Einstein and Maxwell equations have no source terms, the solution-generating
method given by Wald actually allows us to construct the solution for the electromagnetic
test field which occurs when the hole is immersed in an originally uniform magnetic
field surrounded by plasma (i.e., ionized particles) and hence eventually gets 
slightly charged via accretion. We already announced that we shall consider the case when the charge
accreted on the hole is small enough not to distort the background Kerr geometry. Certainly,
it needs to be justified that this is indeed what can actually happen. This will be done
at the end of section II. Since we shall basically employ this algorithm
proposed by Wald, the present work can be thought of as having its basis on the
``first'' approximation of the Einstein-Maxwell system in which only the Maxwell
equation is being solved in the background of Kerr spacetime assuming that neither
the field nor the amount of charge are strong enough to distort the background
geometry. In this sense, any attempt to deal with the problem by directly solving
the coupled Einstein-Maxwell equations, such as the work by Dokuchaev employing
the Ernst-Wild solution to the coupled system, can be regarded as being more 
fundamental although less practical in many respects. Later, we shall notice
that our result for the expression for the magnetic flux through the hole can indeed
be deduced as a leading approximation to that appeared in the work by Dokuchaev
although it has not been realized there.

\begin{center}
{\rm\bf II. Solution Generating Method by Wald}
\end{center}

{\bf 1. Wald field}

From the general properties of Killing fields [7] ; {\it a Killing vector in
a vacuum spacetime generates a solution of Maxwell's equations in the
background of that vacuum spacetime}, long ago, Wald [6] constructed a stationary,
axisymmetric solution of Maxwell's equations in Kerr black hole spacetime. 
To be a little more concrete, Wald's construction is based on the following two
statements : \\
(A) {\it The axial Killing vector $\psi^{\mu} = (\partial/\partial \phi)^{\mu}$
generates a stationary, axisymmetric test electromagnetic field which asymptotically
approaches a uniform magnetic field, has no magnetic monopole moment and has 
charge $= 4J$,} \\
$F_{\psi} = d\psi $ \\
where ``$d$'' denotes the exterior derivative and $J$ is the angular momentum of a 
Kerr black hole. \\
(B) {\it The time translational Killing vector $\xi^{\mu} = (\partial/\partial t)^{\mu}$
generates a stationary, axisymmetric test electromagnetic field which vanishes
asymptotically, has no magnetic monopole moment and has
charge $= -2m$,} \\
$F_{\xi} = d\xi $ \\
where $m$ is the mass of the Kerr hole. \\
Equipped with this preparation, we are now interested in obtaining the solution
for the electromagnetic test field $F$ which occurs when a stationary, axisymmetric
black hole (i.e., Kerr hole) is placed in an originally uniform magnetic field of
strength $B_{0}$ aligned along the symmetry axis of the black hole. And for now
we consider the case when the electric charge is absent. Obviously, then
the solution can be readily written down by referring to the statements (A) and (B) above
as follows ;
\begin{eqnarray}
F = {1\over 2}B_{0} \left[F_{\psi}-\left({4J\over -2m}\right)F_{\xi}\right]
= {1\over 2}B_{0} \left[d\psi + {2J\over m} d\xi \right].
\end{eqnarray}
Then using ; $F={1\over 2}F_{\mu\nu}dx^{\mu}\wedge dx^{\nu}$, 
$\psi = \psi_{\mu}dx^{\mu}$, $\xi = \xi_{\nu}dx^{\nu}$, with
\begin{eqnarray}
\xi_{\mu} &=& g_{\mu\nu}\xi^{\nu}=g_{\mu\nu}\delta^{\nu}_{t}=g_{\mu t},\\
\psi_{\mu} &=& g_{\mu\nu}\psi^{\nu}=g_{\mu\nu}\delta^{\nu}_{\phi}=g_{\mu\phi}
\nonumber
\end{eqnarray}
and for the Kerr metric given in Boyer-Lindquist coordinates [6,14] (or see eq.(61) in
Appendix A), the solution
above can be written in a concrete form as ;
\ba
F=\frac{1}{2}B_{0}[&-&\frac{2ma}{\Sigma^{2}}(r^{2}-a^{2}\cos^{2}\theta)
  (1+\cos^{2}\theta)(dr \wedge dt) \nonumber \\
  &-& \frac{4mra}{\Sigma^{2}}(r^{2}-a^{2})
    \sin\theta \cos\theta (d\theta \wedge dt) \\
  &+& \{ 2r\sin^{2}\theta + \frac{2ma^{2} \sin^{2} \theta}{\Sigma^{2}}(r^{2}-a^{2}\cos^{2}\theta)
    (1+\cos^{2}\theta)\} (dr \wedge d\phi) \nonumber \\
  &+& \{ 2(r^{2}+a^{2})+\frac{4mra^{2}}{\Sigma^{2}}\left[\Sigma \sin^{2}\theta-(r^{2}+a^{2})
    (1+\cos^{2}\theta))\right] \nonumber \\
   && \times \sin\theta \cos\theta \}(d\theta \wedge d\phi)]. \nonumber
\ea
{\bf 2. Wald charge}
\\
We now turn to the issue of {\it charge accretion} onto the Kerr black hole immersed in a 
magnetic field surrounded by an ionized interstellar medium (``plasma''). We shall essentially
follow the argument given in [6] and to do so, we first need to know the physical components
of electric and magnetic fields. This can be achieved by projecting the Maxwell field tensor
given above in eq.(3) onto a tetrad frame. An appropriate tetrad frame here is that suggested
by Carter [6,8] whose physical properties are discussed in detail in the Appendix A.
Now using the
(dual of) Carter's orthonormal tetrad $e_{A} = \{e_{0}, ~e_{1}, ~e_{2}, ~e_{3}\}$ [8],
\begin{eqnarray}
e_{0} &=& \frac{(r^2+a^2)}{(\Sigma \Delta)^{1/2}}\frac{\partial}{\partial t} +
\frac{a}{(\Sigma \Delta)^{1/2}}\frac{\partial}{\partial \phi} = e^{t}_{0}\partial_{t}
+ e^{\phi}_{0}\partial_{\phi}, \nonumber \\
e_{1} &=& \left(\frac{\Delta}{\Sigma}\right)^{1/2}\frac{\partial}{\partial r} = e^{r}_{1}\partial_{r},
\\
e_{2} &=& \frac{1}{\Sigma^{1/2}}\frac{\partial}{\partial \theta} = e^{\theta}_{2}\partial_{\theta},
\nonumber \\
e_{3} &=&  \frac{a\sin \theta}{\Sigma^{1/2}}\frac{\partial}{\partial t} +
\frac{1}{\Sigma^{1/2}\sin \theta}\frac{\partial}{\partial \phi} = e^{t}_{3}\partial_{t}
+ e^{\phi}_{3}\partial_{\phi} \nonumber
\end{eqnarray}
where $\Sigma = r^2+a^2\cos^2 \theta$ and $\Delta = r^2+a^2-2mr$ with $a=J/m$ being the angular
momentum per unit mass and
the Carter tetrad components of the Maxwell field strength can be computed as
\begin{eqnarray}
F &=& B_{0} \left[\frac{ar\sin^{2}\theta}{\Sigma}-\frac{ma}{\Sigma^{2}}(r^{2}-a^{2}\cos^{2}\theta)
   (1+\cos^{2}\theta)\right] (e^{1}\wedge e^{0}) 
   \nonumber \\
 &+& B_{0}\frac{a\Delta^{1/2}}{\Sigma} \sin\theta \cos\theta (e^{2}\wedge e^{0})
  \\
 &+& B_{0}\frac{r\Delta^{1/2}}{\Sigma} \sin\theta (e^{1}\wedge e^{3})
 \nonumber \\
 &+& B_{0}\frac{\cos\theta}{\Sigma}\left[(r^{2}+a^{2})-\frac{2mra^{2}}{\Sigma}  
      (1+\cos^{2}\theta)\right] (e^{2}\wedge e^{3}). \nonumber
\end{eqnarray}
Here, consider particularly the radial component of the electric field (as observed by a local
observer in this Carter tetrad frame) ; 
\begin{eqnarray}
E_{\hat{r}}=E_{1}=F_{10}=B_{0}\left[\frac{ar\sin^{2}\theta}{\Sigma}-\frac{ma}{\Sigma^{2}}
   (r^{2}-a^{2}\cos^{2}\theta)(1+\cos^{2}\theta)\right] \nonumber
\end{eqnarray}
which, along the symmetry axis ($\theta = 0, ~\pi$) of the Kerr hole, becomes
\begin{eqnarray}
E_{\hat{r}}(\theta = 0,~\pi)=-B_{0}\frac{2ma(r^2-a^2)}{\Sigma^2}. \nonumber
\end{eqnarray}
Note that $E_{\hat{r}}(\theta = 0,~\pi)<0$ for $B_{0}>0$ and
$E_{\hat{r}}(\theta = 0,~\pi)>0$ for $B_{0}<0$ meaning that it is radially {\it inward/outward}
if the hole's axis of rotation and the external magnetic field are {\it parallel/antiparallel}.
Put differently, this implies that if the spin of the hole and the magnetic field are {\it parallel},
then {\it positively charged} particles on the symmetry axis of the hole will be
pulled into the hole while if the spin of the hole and the magnetic field are {\it antiparallel},
{\it negatively charged} particles on the symmetry axis of the hole will be
pulled into the hole. In this manner a rotating black hole will ``selectively'' accrete charged
particles until it builds up ``equilibrium'' net charge. Then the next natural question to ask would
be ; how the equilibrium net charge can be determined. To answer this question, we resort to the
``injection energy'' argument originally due to Carter [8].  Recall first that the energy of a charged 
particle in a stationary spacetime with the time translational isometry generated by the Killing 
field $\xi^{\mu}=(\partial /\partial t)^{\mu}$ in the presence of a stationary electromagnetic 
field is given by 
\begin{eqnarray}
\varepsilon = - p_{\alpha}\xi^{\alpha} = - g_{\alpha\beta}p^{\alpha}\xi^{\beta} \nonumber
\end{eqnarray}
with $p^{\mu}={\tilde m}u^{\mu}-eA^{\mu}$ being the 4-momentum of the charged particle with mass
and charge $\tilde{m}$ and $e$ respectively. Now if we lower the charged particle down the
symmetry axis into the Kerr hole, the change in electrostatic energy of the particle will be
\begin{eqnarray}
\delta \varepsilon = \varepsilon_{final} - \varepsilon_{initial} =
eA_{\alpha}\xi^{\alpha}|_{horizon} - eA_{\alpha}\xi^{\alpha}|_{\infty}.
\end{eqnarray}
Now, if $\delta \varepsilon < 0 \rightarrow $ it will be energetically favorable for the hole
to accrete particles with this charge whereas if $\delta \varepsilon > 0 \rightarrow $  
the black hole will accrete particles with opposite charge. In either case, the Kerr hole will
selectively accrete charges until $A^{\mu}$ is changed sufficiently that the electrostatic ``injection
energy'' $\delta \varepsilon$ is reduced to zero. We are then ready to determine, by this injection
energy argument due to Carter, what the equilibrium net charge accreted onto the hole would be.
In the discussion of Wald field given above, we only restricted ourselves to the case of solutions to
Maxwell equation in the background of {\it uncharged} stationary, axisymmetric black hole spacetime
and it was given by eq.(1). Now we need the solution when the stationary, axisymmetric black hole is
slightly charged via charge accretion process described above. Then according to the {\it statement
(B)} in the discussion of Wald field given earlier, there can be at most one more perturbation of a
stationary, axisymmetric vacuum black hole which corresponds to adding a charge $Q$ to the hole and
it is nothing but to linearly superpose the solution $(-Q/2m)F_{\xi} = (-Q/2m)d\xi$ to the solution
given in eq.(1) to get 
\begin{eqnarray}
F = \frac{1}{2}B_{0}[d\psi + \frac{2J}{m}d\xi]-\frac{Q}{2m}d\xi,
\end{eqnarray} 
which, in terms of the gauge potental, amounts to
\begin{eqnarray}
A_{\mu} = {1\over 2}B_{0}(\psi_{\mu}+{2J\over m}\xi_{\mu})-{Q\over 2m}\xi_{\mu}.
\end{eqnarray}
Then the electrostatic injection energy can be computed as
\begin{eqnarray}
\delta \varepsilon &=& eA_{\alpha}\xi^{\alpha}|_{horizon} - eA_{\alpha}\xi^{\alpha}|_{\infty}
\nonumber \\
&=& e\left(\frac{B_{0}J}{m}-\frac{Q}{2m}\right).
\end{eqnarray}
Thus one may conclude that a rotating hole in a uniform magnetic field will accrete charge until
the gauge potential evolves to a value at which $\delta \varepsilon = 0$ yielding the equilibrium
net charge as $Q = 2B_{0}J$. This amount of charge is called ``Wald charge''. \\
Note that we announced from the beginning that we shall consider the case when the charge
accreted on the hole is small enough not to distort the background Kerr geometry. Now
we provide the rationale that this is indeed what can actually happen. To do so, we first
assume that the typical value of the charge on the hole is the equilibrium Wald charge
$Q=2B_{0}J$ just described. Then using the fact that $J=ma \leq m^2$, the charge-to-mass
ratio of a hole has an upper bound
\begin{eqnarray}
{Q \over m} = 2B_{0}\left({J\over m}\right) \leq 2B_{0}m = 2\left({B_{0}\over 10^{15}}\right)
\left({m\over m_{\odot}}\right)10^{-5} \nonumber
\end{eqnarray}
where in the last equality we have converted $B_{0}$ and $m$ from geometrized units to
solar-mass units and gauss [17]. Thus for the two typical examples ; (i) the
binary system in our galaxy with mass $\sim 5\sim 10 ~m_{\odot}$ in the surrounding magnetic
field of strength $\sim 10^{14} (gauss)$, $Q/m \sim 10^{-5} << 1$, and (ii) the AGN (active
galactic nuclei) with mass $\sim 10^6\sim 10^9 ~m_{\odot}$ in the magnetic
field of strength $\sim 10^4 (gauss)$, $Q/m \sim 10^{-7} << 1$. As we can see in these two
cases, the charge-to-mass ratio of the rotating black hole in examples (i) and (ii)
are small enough not to disturb the geometry itself. Thus we can safely
employ the solution-generating method suggested by Wald to construct the solution to
the Maxwell equations when some amounts of charges are around to which we now turn.  

\begin{center}
{\rm\bf III. Stationary, axisymmetric Maxwell field around a ``slightly charged'' Kerr hole}
\end{center}

As discussed in the previous subsection 2, the stationary,
axisymmetric solution to the Maxwell equation in the background of a Kerr hole with charge $Q$
accreted in a originally uniform magnetic field can be constructed as    
\ba
F =\frac{1}{2}B_{0}[F_{\psi}-(\frac{4J}{-2m})F_{\xi}]+(\frac{Q}{-2m})F_{\xi}
=\frac{1}{2}B_{0}[d\psi + \frac{2J}{m}(1-\frac{Q}{2B_{0}J})d\xi]. 
\ea
Again for the Kerr metric given in Boyer-Lindquist coordinates, the solution
above can be written  as ;
\ba
F=\frac{1}{2}B_{0}[&-&\frac{2ma}{\Sigma^{2}}(r^{2}-a^{2}\cos^{2}\theta)
  \{(1-\frac{Q}{B_{0}J})+\cos^{2}\theta\} (dr \wedge dt) 
  \nonumber \\
  &-& \frac{4mra}{\Sigma^{2}}\{r^{2}-a^{2}(1-\frac{Q}{B_{0}J})\} 
    \sin\theta \cos\theta (d\theta \wedge dt) \\
  &+& \{ 2r\sin^{2}\theta + \frac{2ma^{2} \sin^{2} \theta}{\Sigma^{2}}(r^{2}-a^{2}\cos^{2}\theta)
    (1-\frac{Q}{B_{0}J}+\cos^{2}\theta)\} (dr \wedge d\phi) \nonumber \\
  &+& \{ 2(r^{2}+a^{2})+\frac{4mra^{2}}{\Sigma^{2}}\left[\Sigma \sin^{2}\theta-(r^{2}+a^{2})
    (1-\frac{Q}{B_{0}J}+\cos^{2}\theta)\right] \nonumber \\
   && \times \sin\theta \cos\theta \} (d\theta \wedge d\phi)]. \nonumber
\ea
Obviously, in order to have some insight into the nature of this solution to the Maxwell equations,
one may wish to obtain physical components of electric field and magnetic induction. And this can 
only be achieved by projecting the Maxwell field tensor above onto an appropriate tetrad frame
as mentioned earlier.
To be a little more concrete, in Boyer-Lindquist coordinates, one can think of two orthonormal
tetrads, first, the familiar Zero-Angular-Momentum-Observer (ZAMO) tetrad frame [9] and next, 
rather unfamiliar Carter tetrad frame [8]. The ZAMO tetrad is well-known and widely employed in 
various anaysis in the literature. ZAMO is a fiducial observer following timelike geodesic 
orthogonal to spacelike hypersurfaces. And this implies that its 4-velocity is just the timelike
ZAMO tetrad $u^{\mu} = e^{\mu}_{0}$ given below in eq.(12). The Carter tetrad, however, seems
less known than ZAMO despite its physical and technical advantages over ZAMO. Thus we provide
some of the basics of Carter tetrad in the Appendix A. In the following, we just give the components
of Maxwell fields projected on these two tetrad frames without getting into the details of the
nature of the tetrad frames themselves. Upon projecting the Maxwell field tensor components on
a given tetrad frame, the physical components of electric and magnetic fields can be read off
as $E_{i}=F_{i0}=F_{\mu\nu}(e^{\mu}_{i}e^{\nu}_{0})$ and $B_{i}=\epsilon_{ijk}F^{jk}/2=
\epsilon_{ijk}F^{\mu\nu}(e^{\mu}_{j}e^{\nu}_{k})/2$, respectively. \\   
(I) Computation of the ZAMO tetrad components of the Maxwell fields.

From the dual to the ZAMO tetrad, $e^{A}=(e_{0}=e_{(t)}, e_{1}=e_{(r)}, 
e_{2}=e_{(\theta)}, e_{3}=e_{(\phi)})$,
\ba
e_{0} &=& (\frac{A}{\Sigma \Delta})^{1/2}(\partial_{t}+\frac{2mra}{A} \partial_{\phi})
 = e^{t}_{0} \partial_{t} + e^{\phi}_{0} \partial_{\phi}, \nonumber  \\
e_{1} &=& (\frac{\Delta}{\Sigma})^{1/2} \partial_{r}=e^{r}_{1}\partial_{r}, \\
e_{2} &=& \Sigma^{-1/2}\partial_{\theta}=e^{\theta}_{2}\partial_{\theta}, \nonumber  \\
e_{3} &=& (\frac{\Sigma}{A})^{1/2}\frac{1}{\sin\theta}\partial_{\phi}
 =e^{\phi}_{3}\partial_{\phi} \nonumber
\ea
where $A=(r^2+a^2)^2-a^2\Delta \sin^2 \theta$,
we can now read off the ZAMO tetrad components of $F_{\mu \nu}$
\ba
F_{10} &=& F_{\mu\nu} e^{\mu}_{1}e^{\nu}_{0} \nonumber \\
      &=& \frac{B_{0}ma}{(A\Sigma\Delta)^{1/2}\Sigma^{2}}
        [2r^{2}\Sigma^{2}\sin^{2}\theta+(r^{2}-a^{2}\cos^{2}\theta)(1-\frac{Q}{B_{0}J}+\cos^2 \theta )
        \{2mra^{2}\sin^{2}\theta-(\frac{\Delta}{\Sigma})^\frac{1}{2}A\},] \nonumber \\
F_{20} &=& F_{\mu\nu} e^{\mu}_{2}e^{\nu}_{0} \nonumber \\
      &=& \frac{B_{0}mra}{(A\Sigma\Delta)^{1/2}\Sigma^{2}}
        [2(r^{2}+a^{2})\Sigma^{2}+4mra^{2}\{\Sigma \sin^{2}\theta-(r^{2}+a^{2})
        (1-\frac{Q}{B_{0}J}+\cos^{2}\theta)\} \nonumber \\
       && -2\frac{A}{\Sigma^{1/2}}\{r^{2}-a^{2}(1-\frac{Q}{B_{0}J})\}]
       \sin\theta \cos\theta, \nonumber \\
F_{30} &=& F_{\mu\nu} e^{\mu}_{3}e^{\nu}_{0} = 0, \\
F_{12} &=& F_{\mu\nu} e^{\mu}_{1}e^{\nu}_{2} = 0, \nonumber \\
F_{13} &=& F_{\mu\nu} e^{\mu}_{1}e^{\nu}_{3} \nonumber \\
      &=& B_{0}(\frac{\Delta}{A})^{1/2}\frac{\sin\theta}{\Sigma^{2}}
       [(r^{2}+a^{2})r\Sigma+a^{2}\{2r(r^{2}-a^{2})\cos^{2}\theta-(r-m)
       (r^{2}-a^{2}\cos^{2}\theta)(1+\cos^{2}\theta) \nonumber \\
      && -\frac{Qm}{B_{0}J}(r^{2}-a^{2}\cos^{2}\theta)\}], \nonumber \\
F_{23} &=& F_{\mu\nu} e^{\mu}_{2}e^{\nu}_{3} \nonumber \\
      &=& B_{0}\frac{1}{A^{1/2}}\frac{\cos\theta}{\Sigma^{2}}
          [(r^{2}+a^{2})\{(r^{2}-a^{2})(r^{2}-a^{2}\cos^{2}\theta)+2a^{2}r(r-m)(1+\cos^{2}\theta) 
           \nonumber \\
      && +2a^{2}r\frac{Qm}{B_{0}J}\}-a^{2}\Delta\Sigma \sin^{2}\theta]. \nonumber                          
\ea

(II) Computation of the Carter tetrad components of the Maxwell fields.

From the dual to the Carter tetrad, $e_{A}=(e_{0}, e_{1}, e_{2}, e_{3})$, given earlier in eq.(4),
\ba
F_{10} &=& F_{\mu\nu} e^{\mu}_{1}e^{\nu}_{0}
        = B_{0}[\frac{ar \sin^{2}\theta}{\Sigma}-\frac{ma}{\Sigma^{2}}(r^{2}-a^{2}\cos^{2}\theta)
      (1-\frac{Q}{B_{0}J}+\cos^{2}\theta)],   \nonumber \\
F_{20} &=& F_{\mu\nu} e^{\mu}_{2}e^{\nu}_{0}
        = B_{0}\frac{a\Delta^{1/2}}{\Sigma}\sin\theta \cos\theta,  \nonumber \\
F_{30} &=& F_{\mu\nu} e^{\mu}_{3}e^{\nu}_{0} = 0, \\ 
F_{12} &=& F_{\mu\nu} e^{\mu}_{1}e^{\nu}_{2} = 0,  \nonumber \\
F_{13} &=& F_{\mu\nu} e^{\mu}_{1}e^{\nu}_{3} 
        =  B_{0}\frac{r\Delta^{1/2}}{\Sigma}\sin\theta,  \nonumber \\
F_{23} &=& F_{\mu\nu} e^{\mu}_{2}e^{\nu}_{3}
        = B_{0}\frac{\cos\theta}{\Sigma}[(r^{2}+a^{2})-\frac{2mra^{2}}{\Sigma}
      (1-\frac{Q}{B_{0}J}+\cos^{2}\theta)]. \nonumber
\ea 
Therefore, in the Carter tetrad frame,
\ba
F &=& B_{0} [\frac{ar\sin^{2}\theta}{\Sigma}-\frac{ma}{\Sigma^{2}}(r^{2}-a^{2}\cos^{2}\theta)
   (1-\frac{Q}{B_{0}J}+\cos^{2}\theta)] (e^{1}\wedge e^{0}) 
   \nonumber \\
 &+& B_{0}\frac{a\Delta^{1/2}}{\Sigma} \sin\theta \cos\theta (e^{2}\wedge e^{0}) 
  \\
 &+& B_{0}\frac{r\Delta^{1/2}}{\Sigma} \sin\theta (e^{1}\wedge e^{3})  
 \nonumber \\
 &+& B_{0}\frac{\cos\theta}{\Sigma}[(r^{2}+a^{2})-\frac{2mra^{2}}{\Sigma}
      (1-\frac{Q}{B_{0}J}+\cos^{2}\theta)] (e^{2}\wedge e^{3}). \nonumber
\ea
As is pointed out in Appendix A, this Carter tetrad components of the Maxwell field
tensor take much simpler forms than its ZAMO tetrad components given in eq.(13) and
hence are easier to deal with.

\begin{center}
{\rm\bf IV. The magnetic flux across one half of event horizon}
\end{center}

With the asymptotically uniform stationary, axisymmetric magnetic field which is aligned
with the spin axis of a ``slightly charged'' Kerr hole given above, we now would like
to compute the flux of the magnetic field across one half of the surface of the event
horizon which occurs at points where
\be
\Delta(r_{+})=r_{+}^{2}+a^{2}-2mr_{+}=0 \;\;\; {\rm or} \;\;\; r_{+}=m+\sqrt{m^{2}-a^{2}}
\ee
The physical motivation that underlies this study is to have some insight into the question
of how much of the electromagnetic field can actually penetrate into the hole's horizon -
at least in idealized situations. We now begin by considering two vectors lying on the
horizon which, for instance in the Boyer-Lindquist coordinates, are given by
\be
dx_{1}^{\alpha}=(0, \; 0, \; d\theta, \; 0), \;\;\; dx_{2}^{\alpha}=(0, \; 0, \; 0, \; d\phi).
\ee
Then in terms of the 2nd rank tensor constructed from these two vectors [10],
\be
d\sigma^{\alpha\beta}=\frac{1}{2}(dx_{1}^{\alpha}dx_{2}^{\beta}-dx_{1}^{\beta}dx_{2}^{\alpha})
\ee
one now can define the invariant surface element of the horizon as
\ba
ds &=& (2d\sigma_{\alpha\beta}\sigma^{\alpha\beta})^{1/2}\mid_{r_{+}}
   \\
   &=& (g_{\theta\theta}g_{\phi\phi})^{1/2}\mid_{r_{+}} \, d\theta d\phi
   = (r_{+}^{2}+a^{2})\sin\theta d\theta d\phi. \nonumber
\ea
Next, since the tensor $d\sigma^{\alpha\beta}$ is associated with the invariant surface
element of any (not necessarily closed) 2-surface, the flux of electric field and
magnetic field across any 2-surfaces can be given respectively by
\ba
\Phi_{E} &=& \int \tilde{F}_{\alpha\beta} d\sigma^{\alpha\beta} = \int \frac{1}{2}
  \epsilon_{\alpha\beta}^{\;\;\;\;\gamma\delta}F_{\gamma\delta}d\sigma^{\alpha\beta},
  \nonumber \\
\Phi_{B} &=& \int F_{\alpha\beta} d\sigma^{\alpha\beta}.  
\ea
In particular, the flux of magnetic field across one half of the horizon of Kerr hole is
\be
\Phi_{B} = \int_{r=r_{+}} F_{\alpha\beta} d\sigma^{\alpha\beta}
 = \int^{2\pi}_{0}d\phi \int^{\pi/2}_{0} d\theta F_{\theta\phi} \mid_{r_{+}}.
\ee
Then using $F_{\theta \phi}$ evaluated on the horizon
\be
 F_{\theta\phi} \mid_{r_{+}} = B_{0}\frac{\sin\theta \cos\theta}{(r_{+}^{2}+a^{2}\cos^{2}\theta)^{2}}
 (r_{+}^{2}+a^{2})^{2}[r_{+}^{2}-a^{2}(1-\frac{Q}{B_{0}J})]
\ee
and the result of integration
\be
 \int^{2\pi}_{0}d\phi \int^{\pi/2}_{0} d\theta
  \frac{\sin\theta \cos\theta}{(r_{+}^{2}+a^{2}\cos^{2}\theta)^{2}}
 =\frac{\pi}{r_{+}^{2}(r_{+}^{2}+a^{2})}
\ee
we finally get
\be
\Phi_{B}=B_{0}\pi r_{+}^{2}(1+\frac{a^{2}}{r_{+}^{2}})[1-\frac{a^{2}}{r_{+}^{2}}
 (1-\frac{Q}{B_{0}J})].
\ee
Note that thus far we assumed the case when the spin of the hole and the magnetic field
are parallel (i.e., $B_{0}>0$ for $J>0$) and hence the positively charged particles ($Q>0$) are being 
accreted on the hole via Carter's injection energy argument discussed in sect.II
Namely, in calculating the magnetic flux we started with the expression for the solution
$F = (B_{0}/2)[d\psi + (2J/m)d\xi]-(Q/2m)d\xi $ given earlier. If, instead, we consider the 
other case when the hole's rotation axis and the magnetic field are antiparallel (i.e.,
$B_{0}<0$ for $J>0$), the negatively charged particles ($Q<0$) would be accreted on the hole and thus
this time we should start with the solution
$F = (-B_{0}/2)[d\psi + (2J/m)d\xi]+(Q/2m)d\xi $ which has overall sign just {\it opposite}
to that in {\it parallel/positive charge} case. In other words, $B_{0}$ and $Q$ always have the
same sign. This indicates that if we started out with
the {\it antiparallel/negative charge} case, we would end up with the expression for the
magnetic flux having {\it opposite overall} sign to that in eq.(24) above. Therefore, the
general expression for the {\it magnitude} of the magnetic field should take the form
\begin{eqnarray}
\Phi_{B}=|B_{0}|\pi r_{+}^{2}(1+\frac{a^{2}}{r_{+}^{2}})[1-\frac{a^{2}}{r_{+}^{2}}
 (1-\frac{|Q|}{|B_{0}|J})]
\end{eqnarray} 
where now $\Phi_{B}$ is to be understood as denoting the {\it absolute} value of the flux.
Indeed this expression for the magnetic flux through the slightly charged Kerr hole is a
new result that has never been realized in the literature and as such needs close analysis
in some more detail. \\
(i) In the absence of the accretion charge, i.e., $Q=0$, the magnetic flux above correctly
reduces to that obtained long ago by King, Lasota and Kundt [3],
\be
\Phi_{B}^{KLK}=|B_{0}|\pi r_{+}^{2}(1-\frac{a^{4}}{r_{+}^{4}}).
\ee
(ii) With the nonvanishing accretion charge having value in the range $0<|Q|<2|B_{0}|J$,
the total magnetic flux through the hole {\it can never go to zero}. And this property
holds true even when the hole is maximally-rotating, i.e., $a\rightarrow m$,
$r_{+}\rightarrow m$ in which case the total flux becomes 
\be
\Phi_{B}=2|B_{0}|\pi r_{+}^{2}(\frac{|Q|}{|B_{0}|J})=2\pi r_{+}^{2}(\frac{|Q|}{m^{2}})=2\pi |Q|,
\ee
which, interestingly, is independent of $m$, $J$ and $B_{0}$ and has dependence only
on the hole's charge $Q$. Actually, this is the point of central importance we would
like to make in the present work. The physical interpretation of this characteristic
can be briefly stated as follows. When the spin of the hole and the asymtotically
uniform magnetic field are parallel (antiparallel), the hole selectively accretes
positive (negative) charges as we have discussed in the earlier section following the
injection energy argument proposed by Carter and they, in turn, generates magnetic
fields additive to the existing ones. Thus unlike the uncharged Kerr hole case,
the magnetic flux through a slightly charged Kerr hole can never go zero. This effect
manifests itself in a particularly interesting manner when the accreted charge
reaches its maximum value, the Wald charge.  \\
(iii) For the Wald charge value, $|Q|=2|B_{0}|J$,
\be
\Phi_{B}=|B_{0}|\pi r_{+}^{2}(1+\frac{a^{2}}{r_{+}^{2}})^{2}=|B_{0}|4\pi m^{2}
\ee
which is exactly the standard flux across a Schwarzschild black hole. This point is
particularly interesting since the Wald charge, i.e., the amount of equilibrium net
charge accreted on a Kerr hole restores the magnetic flux to the value precisely the
same as that of a uncharged, nonrotating Schwarzschild hole. This point also has been
noticed in the work by Dokuchaev [4] and more recently by Putten [11]. 
Particularly in Dokuchaev's work, it is just this point that led him to conclude that
the magnetic flux through the hole becomes independent of the hole's angular momentum
when the hole has the equilibrium Wald charge. \\
(iv) It is also interesting to note that our result for the magnetic flux through the
slightly-charged Kerr hole given in eq.(24) above can indeed be derived as the leading
approximation to the result obtained from the analysis [4] based on the exact Ernst-Wild
solution [5] to the coupled Einstein-Maxwell equations, although it was not realized
there in the work by Dokuchaev.

\begin{center}
{\rm\bf V. The magnetic flux through a Kerr hole in an oblique configuration case}
\end{center}

Thus far we have considered the case with symmetric geometry in which the stationary,
axisymmetric magnetic field is precisely aligned with Kerr hole's axis of rotation.
It would, however, be of some interest to explore more general case when the asymptotically
uniform, stationary magnetic field happens to be ``oblique'', i.e., aligned at some
angle to the hole's axis of rotation. Then of course the natural question to be addressed
is to see how much of such fields actually can penetrate the horizon and we now turn to
this issue. \\
Indeed the case of uncharged Kerr hole has been studied long ago by Bicak and Janis and
hence in this section, we would like to explore what happens when the Kerr hole is again
``slightly charged'', i.e., $Q \neq 0$, along the same line of analysis as employed
in the previous section. We now statrt with the solution given by Bicak and Janis [10].
The electromagnetic field which is generated when an uncharged Kerr hole is placed in
an originally uniform magnetic field, the direction of which does not coincide with the
hole's axis of rotation has been given by Bicak and Janis and will here be denoted by
$F^{BJ}$ and $A^{BJ}$ for the Maxwell field strength and the associated gauge potential
respectively. They are given in Appendix B.
And in their solution, it is assumed that asymptotically, the field is
decomposed into two components, $B_{0}$ and $B_{1}$, $B_{0}$ being in the direction of
the $z$-axis (i.e., the hole's rotation axis) and $B_{1}$ being chosen to lie,
without loss of generality, along the $x$-axis. Next, since the source-free Maxwell
equation is a linear differential equation, it would admit any linear combination of
particular solutions as another solution. Once again, therefore, we invoke the solution
generating method due to Wald. In particular, in order to construct the solution in
the presence of some charge, we recall that ; according to the {\it statement
(B)} in the discussion of Wald field given earlier, there can be one more perturbation of a
stationary, axisymmetric vacuum black hole which corresponds to adding a charge $Q$ to the hole and
it is nothing but to linearly superpose the solution $(-Q/2m)F_{\xi} = (-Q/2m)d\xi$ to the 
existing solution. Evidently, therefore,
\be
F=F^{BJ}+\frac{-Q}{2m}d\xi \;\;\; {\rm or} \;\;\; A_{\mu}=A_{\mu}^{BJ}+\frac{-Q}{2m}\xi_{\mu}
\ee
constitutes a legitimate solution of Maxwell equations representing the electromagnetic field
around a slightly charged Kerr hole with charge $Q$ which is asymptotically uniform but
is not aligned with the hole's axis of rotation. In fact, this seems to be the only way
available to construct the solution in the presence of some charge and it is worthy of note
that since the particular solution being added, $(-Q/2m)F_{\xi} = (-Q/2m)d\xi$ represents a
axisymmetric test electromagnetic field that {\it vanishes asymptotically}, this new solution
given in eq.(28) above would not change the asymptotic behavior of the Bicak and Janis
solution at all. In Boyer-Lindquist coordinates, thus the
solution we are after can be explicitly written down as
\ba
A_{t} &=& A^{BJ}_{t}+\frac{Q}{2m}\left[\frac{\Delta-a^{2}\sin^{2}\theta}{\Sigma}\right], \nonumber \\
A_{r} &=& A^{BJ}_{r}, \\
A_{\theta} &=& A^{BJ}_{\theta}, \nonumber \\
A_{\phi} &=& A^{BJ}_{\phi}+\frac{Q}{2m}\left[\frac{a\sin^{2}\theta(r^{2}+a^{2}-\Delta)}{\Sigma}\right]
\nonumber 
\ea
for the gauge potential and
\ba
F_{rt} &=& -B_{0}\frac{ma}{\Sigma^{2}}(r^{2}-a^{2}\cos^{2}\theta)(1-\frac{Q}{B_{0}J}+\cos^{2}\theta)
  -B_{1}\frac{mar}{\Sigma^{2}\Delta}\sin\theta \cos\theta \times \nonumber \\
 && [\{r^{3}-2mr^{2}+ra^{2}(1+\sin^{2}\theta)+2ma^{2}\cos^{2}\theta\}\cos\psi
   -a\{r^{2}-4mr+a^{2}(1+\sin^{2}\theta)\}\sin\psi], \nonumber \\
F_{\theta t} &=&  -B_{0}\frac{2mar}{\Sigma^{2}}\sin\theta \cos\theta\{r^{2}-a^{2}(1-\frac{Q}{B_{0}J})\}
       \nonumber \\
       && -B_{1}\frac{ma}{\Sigma^{2}}(r^{2}\cos2\theta+a^{2}\cos^{2}\theta)(a\sin\psi-r\cos\psi),
       \nonumber \\
F_{\phi t} &=&  F_{\phi t}^{BJ}, \nonumber \\
F_{r\theta} &=&  F_{r\theta}^{BJ}, \\
F_{r\phi} &=& B_{0} \frac{\sin^{2}\theta}{\Sigma^{2}}[r\Sigma^{2}+ma^{2}(r^{2}-a^{2}\cos^{2}\theta)
  (1-\frac{Q}{B_{0}J}+\cos^{2}\theta)]\nonumber \\
 && -B_{1} \frac{\sin\theta \cos\theta}{\Delta}
   [(r\Delta-ma^{2})\cos\psi-a(\Delta+mr)\sin\psi] +
   B_{1} \frac{ma^2r}{\Sigma^2 \Delta}\sin^3 \theta \cos \theta \times \nonumber \\
 && [\{r^{3}-2mr^{2}+ra^{2}(1+\sin^{2}\theta)+2ma^{2}\cos^{2}\theta\}\cos\psi
   -a\{r^{2}-4mr+a^{2}(1+\sin^{2}\theta)\}\sin\psi], \nonumber \\
F_{\theta \phi} &=& B_{0} \frac{\sin\theta \cos\theta}{\Sigma^{2}}[(r^{2}+a^{2})
  \{(r^2-a^2)(r^{2}-a^{2}\cos^{2}\theta)+2a^{2}r(r-m)(1+\cos^{2}\theta) \nonumber \\
 && + 2a^{2}r\frac{Qm}{B_{0}J}\}-a^{2}\Delta\Sigma \sin^{2}\theta] \nonumber \\
&& +B_{1} r\sin^{2}\theta[(r-m)\cos\psi - a\sin\psi] \nonumber \\
&& +B_{1}\frac{m\sin^{2}\theta}{\Sigma^{2}}[(r^{2}+a^{2})(r^{2}-a^{2}\cos^{2}\theta)-\Sigma a^2
\cos^2 \theta](r\cos\psi - a\sin\psi) \nonumber 
\ea
for the Maxwell field strength and here $\psi$ has been defined in terms of the azimuthal angle
coordinate $\phi$ as 
\be
\psi=\phi+\frac{a}{r_{+}-r_{-}}ln(\frac{r-r_{+}}{r-r_{-}}).
\ee
This solution in the presence of the accretion charge $Q$ on the Kerr hole reduces to the Bicak-Janis
solution in the absence of the charge given in the Appendix B for $Q=0$ as it should. 
Then in order to calculate the flux of this magnetic field strength
through the horizon of a Kerr hole, we, as usual, need the component $F_{\theta \phi}$ evaluated
on the horizon which is
\ba
F_{\theta \phi}\mid_{r=r_{+}} &=& B_{0} \frac{\sin\theta \cos\theta}{(r_{+}^{2}+a^{2}\cos^{2}\theta)^{2}}
 (r_{+}^{2}+a^{2})^{2}\{r_{+}^{2}-a^{2}(1-\frac{Q}{B_{0}J})\} \nonumber \\
&& +B_{1} r_{+}\sin^{2}\theta \{(r_{+}-m)\cos\phi - a\sin\phi\} \\
&& +B_{1}\frac{m\sin^{2}\theta}{(r_{+}^{2}+a^{2}\cos^{2}\theta)^{2}}
\{(r_{+}^{2}+a^{2})(r_{+}^{2}-2a^{2}\cos^{2}\theta)+a^{4}\sin^{2}\theta \cos^{2}\theta \}
 \nonumber \\
&& \times (r_{+}\cos\phi - a\sin\phi) \nonumber
\ea
where we used
\be
\psi(r_{+})=\phi+\frac{a}{r_{+}-r_{-}} ln(\frac{r_{+}-r_{+}}{r_{+}-r_{-}})=\phi-\infty=\phi.
\ee
We now would like to compute the flux of magnetic field across a generally oriented
one-half portion of Kerr hole's horizon, which shall henceforth be called, ``generally located
hemisphere''. Basically, our goal is the same as before and it is the evaluation of the
magnetic flux across any 2-surface as given by eq.(19) with the invariant surface element of a
Kerr hole's horizon $d\sigma^{\alpha\beta}$ given in eqs.(17) and (18). However, since we now have 
to deal with the invariant magnetic flux across the {\it generally located hemisphere}, we
need a more careful following analysis which is similar to the one encountered when determining
principal axis of spinning rigid body in classical mechanics. Namely, as depicted in Fig.1, we
set a coordinate system in which the hole's axis of rotation coincides with the $z$-axis and
the equatorial plane is represented by $x-y$ plane. Besides, since we are considering the
general case when the asymptotically uniform, stationary magnetic field is ``oblique'', i.e.,
not aligned with the hole's rotation axis, we assume asymptotically and without any loss of
generality that the field can be decomposed into the $z$-component, $B_{0}$ and the $x$-component,
$B_1$. Now, in order to characterize the position of the generally located hemisphere, we
consider rotating the $(x,~y,~z)$-axes by an angle $\beta$ around the $z$-axis and then next
rotating the resulting $(\omega', ~y', ~z)$-axes by an angle $\alpha$ around the $y'$-axis,
to get $(x',~y', ~z')$-axes with now the $z'$-axis representing a kind of ``principal axis''
for the generally located hemisphere. Then the series of coordinate transformations ;
$(x,~y,~z)$ $\rightarrow$ $(\omega', ~y', ~z)$ $\rightarrow$ $(x',~y', ~z')$ obviously
involve two stages of $SO(3)$-rotations and if we denote the spherical-polar angle
coordinates for $(x,~y,~z)$-system and $(x',~y', ~z')$-system by $(\theta, ~\phi)$
and $(\theta', ~\phi')$ respectively, they are related by equations  
\ba
\sin\theta' \cos\phi' &=& \sin\theta  \cos\alpha  \cos(\phi - \beta)-\cos\theta \sin\alpha, 
 \nonumber \\
\sin\theta' \sin\phi' &=& \sin\theta \sin(\phi-\beta), \nonumber \\ 
 \cos\theta' &=& \sin\theta \sin\alpha  \cos(\phi-\beta) + \cos\theta \cos\alpha.
\ea
In the final ``principal axes'' coordinate system, the integration over the polar
angle coordinate $\theta'$ should be done from $0$ to $\pi/2$. Thus we need to determine the 
value of the original polar angle $\theta$ that corresponds to $\theta' = \pi/2$. This can
easily be achieved by plugging $\theta' = \pi/2$ in the last equation above which yields
\be
 \theta=\Theta(\phi; \, \alpha, \, \beta) \equiv \frac{\pi}{2}+tan^{-1}[\tan\alpha
\cos (\phi-\beta)].
\ee
Thus the integration over the generally located hemisphere in the new angle coordinates
$0 \leq \phi' \leq 2\pi$, $0 \leq \theta' \leq \pi/2$ can be translated into that in the
original angle coordinates as
\be
0 \leq \phi \leq 2\pi, \;\;\; 0 \leq \theta \leq \Theta(\phi; \, \alpha, \, \beta)
\ee
and hence finally the magnetic flux across a generally oriented one-half portion of
Kerr hole's horizon reads
\be
\Phi_{B}=\int^{2\pi}_{0}d\phi\int^{\Theta(\phi; \, \alpha, \, \beta)}_{0}d\theta 
 F_{\theta \phi}\mid_{r=r_{+}}.
\ee
Therefore, the magnetic flux across the ``generally located hemisphere'' on 
slightly-charged Kerr hole's horizon is given by
\ba
\Phi_{B} &=& \int^{2\pi}_{0}d\phi\int^{\Theta(\phi; \, \alpha, \, \beta)}_{0}d\theta 
 [F_{\theta \phi}(B_{1}=0)\mid_{r=r_{+}} + F_{\theta \phi}(B_{0}=0)\mid_{r=r_{+}}]
 \nonumber \\
  &=& \Phi(B_{1}=0) + \Phi(B_{0}=0)
\ea
where as given earlier
\ba
F_{\theta \phi}(B_{1}=0)\mid_{r_{+}} &=&  B_{0}\frac{\sin\theta \cos\theta}
 {(r_{+}^{2}+a^{2}\cos^{2}\theta)^{2}}
 (r_{+}^{2}+a^{2})^{2}\{r_{+}^{2}-a^{2}(1-\frac{Q}{B_{0}J})\}, \nonumber \\
F_{\theta \phi}(B_{0}=0)\mid_{r_{+}} &=& B_{1}\frac{\sin^{2}\theta}
 {(r_{+}^{2}+a^{2}\cos^{2}\theta)^{2}}[r_{+}(r_{+}^{2}+a^{2}\cos^{2}\theta)^{2}
 \{(r_{+}-m)\cos\phi-a\sin\phi\} \nonumber \\
&& +m\{(r_{+}^{2}+a^{2})(r_{+}^{2}-2a^{2}\cos^{2}\theta)+a^{4}\sin^{2}\theta \cos^{2}\theta \}
 (r_{+}\cos\phi - a\sin\phi)]. \nonumber
\ea
Just as the case we considered in the preceding section when the asymptotically uniform
magnetic field is aligned with Kerr hole's rotation axis, the effect of adding the Kerr
hole some charge small enough not to disturb its geometry or equivalently adding a
particular solution represented by the Wald field $F=(-Q/2m)d\xi $ changes the value of
$\Phi(B_{1}=0)$ above via the shift in $F_{\theta \phi}(B_{1}=0)\mid_{r_{+}}$ as given.
Since $F_{\theta \phi}(B_{0}=0)\mid_{r_{+}}$ and hence its contribution to the total
flux $\Phi(B_{0}=0)$ remains unchanged upon adding the Wald field, the actual value of
$\Phi(B_{0}=0)$ is precisely as given originally by Bicak and Janis [10]. Thus our task here
simply reduces to the calculation of $\Phi(B_{1}=0)$, namely
\be
 \Phi(B_{1}=0)=B_{0}(r_{+}^{2}+a^{2})^{2}\{r_{+}^{2}-a^{2}(1-\frac{Q}{B_{0}J})\} I_{0}
\ee
with
\be
I_{0} \equiv  \int^{2\pi}_{0}d\phi\int^{\Theta} d\theta \frac{\sin\theta \cos\theta}
 {(r_{+}^{2}+a^{2}\cos^{2}\theta)^{2}}.
\ee
Note, however, this contribution to the total flux across the generally located 
hemisphere is invariant under the rotation about the hole's spin axis (which is
chosen to be the z-axis). Thus in order to go to the principal axis for the
generally located hemisphere, one only needs to perform a single step of $SO(3)$-
rotation $(x,~y,~z)\rightarrow (x',~y',~z')$ which yields the relations between
polar angles
\ba
\sin\theta' \cos\phi' &=& \sin\theta  \cos\alpha  \cos\phi - \cos\theta \sin\alpha, 
 \nonumber \\
\sin\theta' \sin\phi' &=& \sin\theta \sin\phi, \\ 
 \cos\theta' &=& \sin\theta \sin\alpha  \cos\phi + \cos\theta \cos\alpha. \nonumber
\ea
This, in turn, implies that the integration over the generally located hemisphere
actually amounts to the ranges  
\be
0 \leq \phi \leq 2\pi, \;\;\; 0 \leq \theta \leq \Theta(\phi; \, \alpha)
\ee 
with
\be
\Theta(\phi; \, \alpha) \equiv \frac{\pi}{2}+tan^{-1}(\tan\alpha \cos\phi).
\ee
Then the result of actual computation reads
\ba
I_{0} &=& \frac{1}{2r_{+}^{2}(r_{+}^{2}+a^{2})}\int^{2\pi}_{0}d\phi
 \frac{1}{(1+\rho_{+}^{2}\tan^{2}\alpha \cos^{2}\phi)} \nonumber \\
 &=&  
 \frac{\pi}{r_{+}^{2}(r_{+}^{2}+a^{2})}(1+\rho_{+}^{2}\tan^{2}\alpha)^{-1/2}
\ea
and hence
\be
 \Phi(B_{1}=0)=B_{0}\pi 
 r_{+}^{2}(1+\frac{a^{2}}{r_{+}^{2}})\{1-\frac{a^{2}}{r_{+}^{2}}
 (1-\frac{Q}{B_{0}J})\}(1+\rho_{+}^{2}\tan^{2}\alpha)^{-1/2}
\ee
where $\rho_{+}^{2} \equiv (1+a^{2}/r_{+}^{2}).$
Finally, putting this result of ours, $\Phi(B_{1}=0)$ together with the contribution
coming from the $B_{1}$-component of the asymptotically uniform magnetic field,
$\Phi(B_{0}=0)$, we arrive at the total magnetic flux across the generally located
hemisphere on slightly-charged Kerr hole's horizon
\begin{eqnarray}
\Phi_{B} &=& \Phi(B_{1}=0) + \Phi(B_{0}=0) \\
&=& B_{0}\pi  r_{+}^{2}(1+\frac{a^{2}}{r_{+}^{2}})\{1-\frac{a^{2}}{r_{+}^{2}}
 (1-\frac{Q}{B_{0}J})\}(1+\rho_{+}^{2}\tan^{2}\alpha)^{-1/2} \nonumber \\
&+& 2B_{1}\left[r_{+}\{(r_{+}-m)\cos \beta - a\sin \beta\}I_{1} + m
(r_{+}\cos \beta - a\sin \beta)(2\rho_{+}\bar{I_{1}}-I_{1}-a^2I_{2})\right]
\nonumber
\end{eqnarray}
where
\ba
I_{1} &=& \frac{\pi}{2} \tan\alpha (1+\tan^{2}\alpha)^{-1/2}, \nonumber \\
\bar{I_{1}} &=& \frac{\pi}{2} \rho_{+} \tan\alpha 
(1+\rho_{+}^{2}\tan^{2}\alpha)^{-1/2}, \\
I_{2} &=& \frac{\pi}{a^{2}\tan\alpha}\{(1+\rho_{+}^{2}\tan^{2}\alpha)^{1/2}
  - (1+\tan^{2}\alpha)^{1/2}\}. \nonumber
\ea
Finally, some discussions on interesting observations are in order. \\
(i) In the absence of the accretion charge, i.e., $Q=0$, the total magnetic flux above 
correctly reduces to that obtained by Bicak and Janis as it should. \\
(ii) For $\alpha = 0$, namely over the hemisphere which is symmetrically located
around the hole's rotation axis, we have
\be
\Phi_{B}=|B_{0}|\pi r_{+}^{2}(1+\frac{a^{2}}{r_{+}^{2}})[1-\frac{a^{2}}{r_{+}^{2}}
 (1-\frac{|Q|}{|B_{0}|J})] \nonumber
\ee
namely, one recovers the result in eq.(25) for the case where the asymptotically uniform
magnetic field is aligned with the hole's spin axis we studied earlier. Obviously
this was expected since with this orientation of the hemisphere, the contribution
to the total flux coming from the $B_{1}$-component (which is perpendicular to
the hole's spin axis) is zero. Here, the points worthy of note are essentially
the same as before. First, with the nonvanishing accretion charge $Q \neq 0$, 
the total magnetic flux through the hole {\it can never go to zero}. 
In particular, when the hole is maximally-rotating, the total flux becomes
$\Phi_{B}=2\pi |Q|$ which is independent of $m$, $J$ and $B_{0}$ and depends only
on the hole's charge. Lastly, when the accreted charge takes the particular value, 
$|Q|=2|B_{0}|J$, the total flux gets maximized and it is precisely the 
standard flux across a Schwarzschild black hole, $\Phi_{B}=|B_{0}|4\pi m^{2}$
as pointed out earlier. \\
(iii) For $\alpha = \pi/2$, namely when the principal axis for the hemisphere
is perpendicular to the hole's rotation axis, we have
\begin{eqnarray}
\Phi_{B} = B_{1}\pi [r_{+}^2\cos \beta - (r_{+}+m)a\sin \beta ].
\end{eqnarray}
The fact that the total flux depends only on the $B_{1}$-component is also
expected since with this orientation of the hemisphere, the $B_{0}$-component
contribution to the total flux is obviously zero. Next, the angle $\beta$
determines, asymptotically, the angle between the principal axis for the
hemisphere and the axis $x$ which is along the $B_{1}$-component of the field.
As such, the total flux gets maximum for $\beta = 0$ when
$\Phi_{B} = B_{1}\pi r_{+}^2$ and gets minimum for $\beta = \beta_{0}$, with
$\tan \beta_{0} = [r_{+}^2/a(r_{+}+m)]$, when $\Phi_{B} = 0$. 
For extreme Kerr hole, $a = m = r_{+}$ and hence $\tan \beta_{0} = 1/2$ or
$\beta_{0} = 27^{o}$. Moreover, since we may assume $-\pi/2 \leq \beta \leq
\pi/2$, $\cos \beta_{0}$ is always positive and hence  $\Phi_{B} = 0$ occurs
only if $a\sin \beta_{0}>0$. Thus if $a>0$, then $\beta_{0}>0$ and this confirms
our intuition that the field lines are bent near the hole in the same direction
in which the hole rotates since the rotating hole drags field lines along. \\
(iv) Now in this case when the rotating hole takes small amount of accretion
charge, one might wonder how come only the part of the magnetic flux 
$\Phi(B_{1}=0)$ coming from the $B_{0}$-component of the field gets affected with the
other part $\Phi(B_{0}=0)$ coming from the $B_{1}$-component of the field remaining unchanged.
In fact, it has been anticipated from the way we constructed the solution to the
Maxwell equations in the presence of some charge. Namely, notice that we employed
the solution generating scheme by Wald in which the particular solution
$(-Q/2m)F_{\xi} = (-Q/2m)d\xi$ has been superposed to another particular solution,
i.e., the Bicak-Janis solution $F^{BJ}$. And this solution, $F = (-Q/2m)d\xi$ represents
axisymmetric electromagnetic test field aligned precisely with the hole's rotation
axis. Besides, this particular solution vanishes asymptotically not affecting the
asymptotic behavior of the Bicak-Janis solution. Therefore, only the $B_{0}$-component
of the Bicak-Janis solution has been modified locally and thus the $\Phi(B_{1}=0)$
part of the magnetic flux gets affected as a consequence. \\
(v) Lastly, we point out that in this generally oblique geometry case, the charge
on the hole $Q$ has been regarded as being arbitrary. Indeed, the application
of Carter's injection energy argument to the determination of equilibrium charge
$Q=2B_{0}J$ we discussed earlier was quite straightforward for the case when the 
field and the hole's
spin are exactly aligned. And of course it can be attributed to the simple structure
of the solution to the Maxwell equations given in terms of the exterior derivatives
of Killing fields as can be seen in eqs.(7) and (8). This kind of advantage, however,
does not seem to be available in the present oblique geometry case as the solution
given in eq.(30) or (31) no longer possesses such a previliged structure. As a result,
the determination of the equilibrium charge value is unlikely to be successful here
although such equilibrium charge is still expected to exist in principle.

\begin{center}
{\rm\bf VI. Concluding remarks}
\end{center}
  
In the present work, based on the solution-generating method given by Wald, it has 
been demonstrated in a transparent manner
that in a more realistic situation when there are charges around (but small enough
not to disturb the background vacuum geometry), the magnetic flux penetrating the horizon
depends not only on the angular momentum but on the amount of charge accreted on the hole
as well and can never vanish no matter how fast the hole spins. The points worthy of note 
can be summarized as follows. First we start with the case when the asymptotically
uniform magnetic field and the hole's spin are precisely aligned with each other.
Then by the argument given by Wald concerning the charge accretion process, the hole
will gradually accretes the charge until it reaches the equilibrium value $|Q|=2|B_{0}|J$.
Thus with the nonvanishing accretion charge having value in
the range $0 < |Q| < 2|B_{0}|J$, the total magnetic flux through the hole {\it can never go to zero}.
In particular, when the hole is maximally-rotating, the total flux becomes
$\Phi_{B}=2\pi |Q|$ which is independent of $m$, $J$ and $B_{0}$ and depends only  
on the hole's charge. Lastly, when the accreted charge reaches its maximum value,
$|Q|=2|B_{0}|J$, the total flux also gets maximized and it is precisely the
standard flux across a Schwarzschild black hole, $\Phi_{B}=|B_{0}|4\pi m^{2}$
as pointed out earlier. Next, the physical interpretation of this characteristic
can be briefly stated as follows. When the spin of the hole and the asymtotically
uniform magnetic field are parallel (antiparallel), the hole selectively accretes
positive (negative) charges as we have discussed in the earlier section following the
injection energy argument proposed by Carter and they, in turn, generates magnetic
fields additive to the existing ones. Thus unlike the uncharged Kerr hole case,
the magnetic flux through a slightly charged Kerr hole can never go zero. 
We have mainly considered the case with symmetric geometry in which the stationary,
axisymmetric magnetic field is precisely aligned with Kerr hole's axis of rotation.
It would, however, be of some interest to explore more general case when the asymptotically
uniform, stationary magnetic field happens to be ``oblique'', i.e., aligned at some
angle to the hole's axis of rotation. Then of course the natural question to be addressed
is to see how much of such fields actually can penetrate the horizon. Thus we also
explored what would happen when the Kerr hole is again ``slightly charged''
along the same line of analysis as the one employed in the previous 
symmetric geometry case. In this oblique geometry case, however, although we could
write down the solution to the Maxwell equations and evaluate the magnetic flux
through the hole in a quite straightforward manner using basically the solution 
generating scheme by Wald, the determination of equilibrium charge value was not 
attempted due to technical barriers. This issue might be worth persuing and we hope
we can come back to it in the future work. It is our belief that the result of this work, 
obtained in a more realistic situation when some amount of charges are around, 
eventually lends supports to the operational nature of the Blanford-Znajek mechanism 
and puts it on a firmer ground at least on the theoretical side.

\vspace*{1cm}

\begin{center}
{\rm\bf Acknowledgements}
\end{center}

This work was supported in part by Brain Korea 21 project. CHL and HKL are also supported 
in part by the interdisciplinary research program of the KOSEF, Grant No. 1999-2-003-5.

\vspace*{2cm}

\begin{center}
{\rm\bf Appendix A : The Carter tetrad}
\end{center}

Generally speaking, in order to represent a given background geometry, one needs to first 
choose a coordinate system in which the metric is to be given and next, in order to obtain 
physical components of a tensor (such as the electric and magnetic field values), one has to 
select a tetrad frame (in a given coordinate system) to which the tensor components are to be
projected. For the Kerr background spacetime, here we choose the Boyer-Lindquist coordinates 
which can be viewed as the generalization of Schwarzschild coordinates to the stationary,
axisymmetric case. Turning to the choice of tetrad frame, there are
largely two types of tetrad frames ; orthonormal tetrad and null tetrad. As is well-known,
the orthonormal tetrad is a set of four mutually orthogonal unit vectors at each
point in a given spacetime which give the directions of the four axes of locally-
Minkowskian coordinate system
\begin{eqnarray}
ds^2 &=& g_{\mu\nu}dx^{\mu}dx^{\nu} = \eta_{AB}e^{A}e^{B} \nonumber \\
&=& -(e^0)^2 + (e^1)^2 + (e^2)^2 + (e^3)^2 
\end{eqnarray}
where $e^{A}=e^{A}_{\mu}dx^{\mu}$. Namely, every physical observer with 4-velocity
$u^{\mu}$ has associated with him an orthonormal frame in which the basis vectors are
the (reciprocal of) orthonormal tetrad $e_{A}=\{e_{0}=u, ~e_{1}, ~e_{2}, ~e_{3}\}$.
And corresponding to this is a null tetrad $Z_{A}=\{l, ~n, ~m, ~\bar{m}\}$
defined by
\begin{eqnarray}
e_{0} &=& {1\over \sqrt{2}}(l + m), ~~~e_{1} = {1\over \sqrt{2}}(l - m), \\
e_{2} &=& {1\over \sqrt{2}}(m + \bar{m}), ~~~e_{2} = {1\over \sqrt{2}i}(m - \bar{m})
\nonumber
\end{eqnarray}
satisfying the orthogonality relation
\begin{eqnarray}
-l^{\mu}n_{\mu} = 1 = m^{\mu}\bar{m}_{\mu}
\end{eqnarray}
with all other contractions being zero and
\begin{eqnarray}
g^{\mu\nu} = - l^{\mu}n^{\nu} - n^{\mu}l^{\nu} + m^{\mu}\bar{m}^{\nu}
+ \bar{m}^{\mu}m^{\nu}.
\end{eqnarray}
Conversely, given a non-singular null tetrad, there is a corresponding physical
observer. The tetrad vectors then can be used to obtain, from tensors in
arbitrary coordinate system, their physical (i.e., finite and non-zero) components
measured by an observer in this locally-flat tetrad frame. And the rules for
calculating the physical components of a tensor, say, $T_{\mu\nu}$ in the orthonormal
frame and in the null frame are given respectively by
\begin{eqnarray}
T_{AB} = T_{\mu\nu}(e_{A}^{\mu}e_{B}^{\nu}),
~~~T_{lm} = T_{\mu\nu}(l^{\mu}m^{\nu}), ~~~{\rm etc.}
\end{eqnarray}
where $e_{A}^{\mu}$ is the inverse of the tetrad vectors $e^{A}_{\mu}$ in that
$e_{A}^{\mu}e^{A}_{\nu} = \delta^{\mu}_{\nu}$ and $e_{A}^{\mu}e^{B}_{\mu} = \delta_{A}^{B}$.
As just stated, all that is required of the ``correct'' boundary
conditions for electric and magnetic fields at the horizon is that
the physical field's components in the neighborhood of an event horizon should have
``nonspecial'' values. Or put another way, a physically well-behaved observer at the 
horizon should see the fields as having finite and non-zero values. One such choice
of well-behaved tetrad frame has been suggested long ago by Carter [8].
The construction of Carter's orthonormal tetrad starts from Kinnersley's null tetrad [12].
In Boyer-Lindquist coordinates $x^{\mu}=(t,r,\theta,\tilde{\phi})$, its contravariant
and covariant components are given by 
\begin{eqnarray}
l^{\mu} &=& \left({(r^2+a^2)\over \Delta}, ~~1, ~~0, ~~{a\over \Delta}\right), \nonumber \\
n^{\mu} &=& \left({(r^2+a^2)\over 2\Sigma}, ~~{-\Delta \over 2\Sigma}, ~~0, 
~~{a\over 2\Sigma}\right), \\
m^{\mu} &=& {1\over \sqrt{2}(r+ia \cos \theta)}\left(ia\sin \theta, ~~0, ~~1, 
~~{i\over \sin \theta}\right) \nonumber
\end{eqnarray}
and
\begin{eqnarray}
l_{\mu} &=& g_{\mu\nu}l^{\nu} = \left(1, ~~{-\Sigma \over \Delta}, ~~0, ~~a\sin^2 \theta \right), 
\nonumber \\
n_{\mu} &=& g_{\mu\nu}n^{\nu} = \left({\Delta \over 2\Sigma}, ~~{1\over 2}, ~~0,
~~{-a\Delta \over 2\Sigma}\sin^2 \theta \right), \\
m_{\mu} &=& g_{\mu\nu}m^{\nu} = {1\over \sqrt{2}(r+ia \cos \theta)}\left(ia\sin \theta, ~~0, 
~~-\Sigma, ~~-i(r^2+a^2)\sin \theta \right). \nonumber
\end{eqnarray}
where, as before, $\Sigma = r^2+a^2\cos^2\theta $ and $\Delta = r^2+a^2-2Mr $ with $M$ and $a$
being the ADM mass and the angular momentum per unit mass of the hole respectively.
This Kinnersley's null tetrad has been chosen so that $l^{\mu}$ and $n^{\mu}$ lie along the
two principal repeated null directions of the Weyl tensor. Kinnersley's null tetrad has proved
very useful for separating and solving the equations governing scalar, electromagnetic and
gravitational perturbations of Kerr geometry [13]. However, the associated orthonormal tetrad
suffers from two disadvantages. It is singular on the horizon and an observer at rest in it
has non-zero radial velocity. This last point is caused by the asymmetric normalization of
$l^{\mu}$ and $n^{\mu}$, and means that the corresponding orthonormal tetrad is unnatural
in that it frequently hides interesting features of the fields. For these reasons and others,
one obtains another null tetrad and the associated orthonormal tetrad (Carter tetrad) by
``null rotating'' the Kinnersley's null tetrad. Thus at this point, it seems relevant to recall 
some of the basics of null rotation.
Notice that the orthogonality relations for null tetrad given in eq.(53) remain invariant under the
6-parameter group of homogeneous Lorentz transformations at each point of spacetime.
And this Lorentz group can be decomposed into 3-Abelian subgroups ;
\begin{eqnarray}
(I) ~~l&\rightarrow& l, ~~~m\rightarrow m+al,
    ~~~n\rightarrow n+a\bar{m}+\bar{a}m+a\bar{a}l, \nonumber \\
(II) ~~n&\rightarrow& n, ~~~m\rightarrow m+bn, 
     ~~~l\rightarrow l+b\bar{m}+\bar{b}m+b\bar{b}n, \\
(III) ~~l&\rightarrow& \Lambda l, ~~~n\rightarrow \Lambda^{-1} n, 
      ~~~m\rightarrow e^{i\theta}m \nonumber
\end{eqnarray}
where $a$ and $b$ are complex numbers and $\Lambda$ and $\theta$ are real.
Each of these group transformations are called a ``null rotation'' [14] and here
we particularly consider the null rotation (III). Under this null rotation (III),
the corresponding orthonormal tetrad $e_{A}$ is boosted in the $e_{1}=e_{\hat{r}}$
direction with 3-velocity ${(\Lambda^2-1)/(\Lambda^2+1)}$ and spatially rotated
about $e_{1}=e_{\hat{r}}$ through the angle $\theta$. Indeed this action is 
precisely what we need. Namely, in order to get a null tetrad well-behaved at
the horizon, we need to boost it by an amount that becomes suitably infinite
on the horizon. Thus we perform the null rotation (III) on the Kinnersley's
null tetrad with 
$\Lambda = (\Delta/2\Sigma)^{1/2}$ and $e^{i\theta}=\Sigma^{1/2}/(r-ia\cos\theta)$
to obtain the following non-singular null tetrad on the horizon ;
\begin{eqnarray}
l^{'\mu} &=& \left({(r^2+a^2)\over (2\Sigma\Delta)^{1/2}}, ~~({\Delta \over 2\Sigma})^{1/2}, 
~~0, ~~{a\over (2\Sigma\Delta)^{1/2}}\right), \nonumber \\
n^{'\mu} &=& \left({(r^2+a^2)\over (2\Sigma\Delta)^{1/2}}, ~~-({\Delta \over 2\Sigma})^{1/2}, 
~~0, ~~{a\over (2\Sigma\Delta)^{1/2}}\right), \\
m^{'\mu} &=& {1\over \sqrt{2}\Sigma^{1/2}}\left(ia\sin \theta, ~~0, ~~1,
~~{i\over \sin \theta}\right). \nonumber
\end{eqnarray}
Then the associated orthonormal tetrad is
\begin{eqnarray}
e^{\mu}_{0} &=& \left({(r^2+a^2)\over (\Sigma\Delta)^{1/2}}, ~~0,
~~0, ~~{a\over (\Sigma\Delta)^{1/2}}\right), \nonumber \\
e^{\mu}_{1} &=& \left(0, ~~({\Delta \over \Sigma})^{1/2}, ~~0, ~~0\right), \\
e^{\mu}_{2} &=& \left(0, ~~0, ~~{1\over \Sigma^{1/2}}, ~~0\right) \nonumber \\
e^{\mu}_{3} &=& \left({a\sin \theta \over \Sigma^{1/2}},
~~0, ~~0, ~~{1\over \Sigma^{1/2}\sin \theta}\right) \nonumber
\end{eqnarray}
and its dual is given by Carter as
\begin{eqnarray}
e^{0}_{\mu} &=& \left(({\Delta \over \Sigma})^{1/2}, ~~0, ~~0,
~~-({\Delta \over \Sigma})^{1/2} a\sin^2 \theta \right), \nonumber \\
e^{1}_{\mu} &=& \left(0, ~~({\Sigma \over \Delta})^{1/2}, ~~0, ~~0\right), \\  
e^{2}_{\mu} &=& \left(0, ~~0, ~~\Sigma^{1/2}, ~~0\right) \nonumber \\
e^{3}_{\mu} &=& \left({-a\sin \theta \over \Sigma^{1/2}},
~~0, ~~0, ~~{(r^2+a^2)\over \Sigma^{1/2}}\sin \theta \right). \nonumber
\end{eqnarray}
Now, in order to have some insight into the nature of this Carter's orthonormal tetrad
in Boyer-Lindquist coordinates, we first rewrite the metric of Kerr geometry as implied
by this Carter's dual tetrad ;
\begin{eqnarray}
ds^2 &=& \eta_{AB}e^{A}e^{B} \\
&=& -{\Delta\over \Sigma}[dt - a\sin^2 \theta d\phi]^2 + {\Sigma \over \Delta}dr^2
+ \Sigma d\theta^2 + {\sin^2 \theta \over \Sigma}[(r^2+a^2)d\phi - adt]^2. \nonumber
\end{eqnarray}
Then one can immediately realize that an observer at rest in this Carter frame travels
around the hole at ($r=$const., $\theta=$const.) with the angular velocity,
$\Omega^{c}=a/(r^2+a^2)$ which is independent of $\theta$. Certainly, this is in contrast
to what happens in ZAMO (or LNRF) tetrad frame in which case a ZAMO observer travels
around the hole with the same angular velocity as that of the hole,
$\Omega = a[(r^2+a^2)-\Delta]/[(r^2+a^2)^2-\Delta a^2\sin^2 \theta ]$ which has dependence
on the polar angle $\theta$. Indeed, the physical significance of the Carter tetrad frame
is that observers at rest in it see principal null congruence photons moving with purely
radial velocities. This leads one to suppose that Maxwell equations and their solutions
should take relatively simple forms in this Carter tetrad frame as first pointed out by
Znajek [16] and it is indeed the case.
\\
\begin{center}
{\rm\bf Appendix B : The Bicak-Janis solution}
\end{center}
Here, we provide the electromagnetic field which is generated when a Kerr black hole is
placed in an originally uniform magnetic field with its direction not coinciding with
the rotation axis of the hole. This solution has been given by Bicak and Janis [10].
As is well-known, in Newman-Penrose formalism [14], the three independent complex null tetrad
components of the Maxwell field strength tensor are given by
\begin{eqnarray}
\phi_{0} &=& F_{\mu\nu}l^{\mu}m^{\nu}, \nonumber \\
\phi_{1} &=& {1\over 2}F_{\mu\nu}(l^{\mu}n^{\nu}+\bar{m}^{\mu}m^{\nu}), \\
\phi_{2} &=& F_{\mu\nu}\bar{m}^{\mu}n^{\nu}. \nonumber
\end{eqnarray}
The explicit form of $\phi_{0}$, $\phi_{1}$ and $\phi_{2}$ as a solution of Maxwell 
equations is given in the earlier work of Bicak and Dvorak [15] and it contains parameters
$B^{x}_{0}$, $B^{y}_{0}$ and $B^{z}_{0}$ denoting the components of the oblique magnetic
field in asymptotically Minkowskian coordinates $x=r\sin \theta \cos \phi$,
$y=r\sin \theta \sin \phi$, $z=r\cos \theta$. Without any loss of generality, however,
one can put $B^{y}_{0}=0$ and denote by $B^{x}_{0}\equiv B_{1}$, the field component
perpendicular to the rotation axis and by $B^{z}_{0}\equiv B_{0}$, the field component
aligned along the rotation axis. Now given the solutions ${\phi_{0}, ~\phi_{1}, ~\phi_{2}}$,
we would like to have the expression in terms of Maxwell field strength $F_{\mu\nu}$, say,
in Boyer-Lindquist coordinates. This can be achieved first by inverting the above expression
\begin{eqnarray}
F_{\mu\nu} &=& 2(\phi_{1}+\bar{\phi}_{1})n_{[\mu}l_{\nu]} + 2(\phi_{1}-\bar{\phi}_{1})
m_{[\mu}\bar{m}_{\nu]} \\
&+& 2\phi_{2}l_{[\mu}m_{\nu]} + 2\bar{\phi}_{2}l_{[\mu}\bar{m}_{\nu]} + 
2\phi_{0}\bar{m}_{[\mu}n_{\nu]} + 2\bar{\phi}_{0}m_{[\mu}n_{\nu]} \nonumber
\end{eqnarray}
and evaluating this with the standard Kinnersley's null tetrad (i.e., the covariant components)
given earlier in Appendix A. The result is
\begin{eqnarray}
F_{rt} &=& -B_{0}\frac{ma}{\Sigma^{2}}(r^{2}-a^{2}\cos^{2}\theta)(1+\cos^{2}\theta)
  -B_{1}\frac{mar}{\Sigma^{2}\Delta}\sin\theta \cos\theta \times \nonumber \\
 && [\{r^{3}-2mr^{2}+ra^{2}(1+\sin^{2}\theta)+2ma^{2}\cos^{2}\theta\}\cos\psi
   -a\{r^{2}-4mr+a^{2}(1+\sin^{2}\theta)\}\sin\psi], \nonumber \\
F_{\theta t} &=&  -B_{0}\frac{2mar}{\Sigma^{2}}\sin\theta \cos\theta (r^{2}-a^{2})
       \nonumber \\
       && -B_{1}\frac{ma}{\Sigma^{2}}(r^{2}\cos2\theta+a^{2}\cos^{2}\theta)(a\sin\psi-r\cos\psi),
       \nonumber \\
F_{\phi t} &=& -B_{1}\frac{ma}{\Sigma}\sin \theta \cos \theta (r\sin \psi + a\cos \psi), \nonumber \\
F_{r\theta} &=& -B_{1}\frac{1}{\Delta}[\Delta (r\sin \psi + a\cos \psi) \nonumber \\
  && + a\{(mr-a^2\sin^2 \theta)\cos \psi - a(r\sin^2 \theta + m\cos^2 \theta)\sin \psi \}], \\
F_{r\phi} &=& B_{0} r\sin^{2}\theta \nonumber \\
 && -B_{1} \frac{\sin\theta \cos\theta}{\Delta}
   [(r\Delta-ma^{2})\cos\psi-a(\Delta+mr)\sin\psi] - a\sin^2 \theta F_{rt}, \nonumber \\
F_{\theta \phi} &=& B_{0} \Delta \sin\theta \cos\theta \nonumber \\
&& +B_{1}[(r^2 \sin^2 \theta + mr\cos 2\theta)\cos\psi - a(r\sin^2 \theta + m\cos^2 \theta)\sin\psi] 
- \frac{(r^2+a^2)}{a}F_{\theta t} \nonumber
\end{eqnarray}
for the Maxwell field strength tensor and 
\begin{eqnarray}
A_{t} &=& B_{0}\frac{a}{\Sigma}[-\Sigma + mr(1+\cos^2 \theta)] + B_{1}\frac{ma}{\Sigma}\sin \theta
\cos \theta (r\cos \psi - a\sin \psi), \nonumber \\
A_{r} &=& -B_{1}(r-m)\sin \theta \cos \theta \sin \psi, \\
A_{\theta} &=&  -B_{1}[a(r\sin^2 \theta + m\cos^2 \theta)\cos \psi + 
(r^2 \cos^2 \theta - mr\cos 2\theta + a^2\cos 2\theta)\sin \psi ], \nonumber \\
A_{\phi} &=& B_{0}\frac{\sin^2 \theta}{2\Sigma}[\Sigma (r^2+a^2) - 2a^2mr(1+\cos^2 \theta)] \nonumber \\
&& - B_{1}\frac{\sin \theta \cos \theta}{\Sigma}[\Sigma \Delta \cos \psi + m(r^2+a^2)(r\cos \psi -
a\sin \psi)] \nonumber
\end{eqnarray}
for the corresponding gauge potential. To summarize, this is the solution in the absence of the charge
on the hole which is to be compared with the solution given in the text in the presence of the accretion
charge on the Kerr black hole $Q$.

\noindent

\begin{center}
{\rm\bf References}
\end{center}

\begin{description}

\item {[1]} R. D. Blanford and R. L. Znajek, Mon. Not. R. Astron. Soc. {\bf 179}, 433 (1977) ;
            as good review articles, also see, D. Macdonald and K. S. Thorne, {\it ibid} {\bf 198},
            345 (1982) ; H. K. Lee, R. A. M. J. Wijers, and G. E. Brown, Phys. Rep. {\bf 325},
            83 (2000). 
\item {[2]} M. J. Rees, M. C. Begelman, R. D. Blanford, and E. S. Phinney, Nature {\bf 295},
            17 (1982).
\item {[3]} A. R. King, J. P. Lasota, and W. Kundt, Phys. Rev. {\bf D12}, 3037 (1975).
\item {[4]} V. I. Dokuchaev, Sov. Phys. JETP {\bf 65}, 1079 (1987).
\item {[5]} F. J. Ernst and W. J. Wild, J. Math. Phys. {\bf 17}, 182 (1976).
\item {[6]} R. M. Wald, Phys. Rev. {\bf D10}, 1680 (1974).
\item {[7]} A. Papapetrou, Ann. Inst. H. Poincare {\bf 4}, 83 (1966).
\item {[8]} B. Carter, in {\it Black holes}, eds C. DeWitt and B. DeWitt, Gordon and Breach, 
            New York (1973).
\item {[9]} J. M. Bardeen, Astrophys. J. {\bf 162}, 71 (1970).
\item {[10]} J. Bicak and V. Janis, Mon. Not. R. Astron. Soc. {\bf 212}, 899 (1985).
\item {[11]} M. H. P. M. van Putten, Phys. Rev. Lett. {\bf 84}, 3752 (2000).
\item {[12]} W. Kinnersley,  J. Math. Phys. {\bf 10}, 1195 (1969).
\item {[13]} S. A. Teukolsky, Astrophys. J. {\bf 185}, 635 (1973) ; W. H. Press and
             S. A. Teukolsky, {\it ibid}, {\bf 185}, 649 (1973) ; S. A. Teukolsky and
             W. H. Press, {\it ibid}, {\bf 193}, 443 (1974).
\item {[14]} See, for instance, S. Chandrasekhar, {\it The Mathmatical Theory of Black Holes},
             Oxford Univ. Press (1992).
\item {[15]} J. Bicak and L. Dvorak, Gen. Rel. Grav. {\bf 7}, 959 (1976).
\item {[16]} R. L. Znajek, Mon. Not. R. Astron. Soc. {\bf 179}, 457 (1977).
\item {[17]} H. K. Lee, C. H. Lee, and M. H. P. M. van Putten, {\it astro-ph/0009239}.

\end{description}

\newpage 

{\bf Figure Caption}
\\
\\
\\
\\
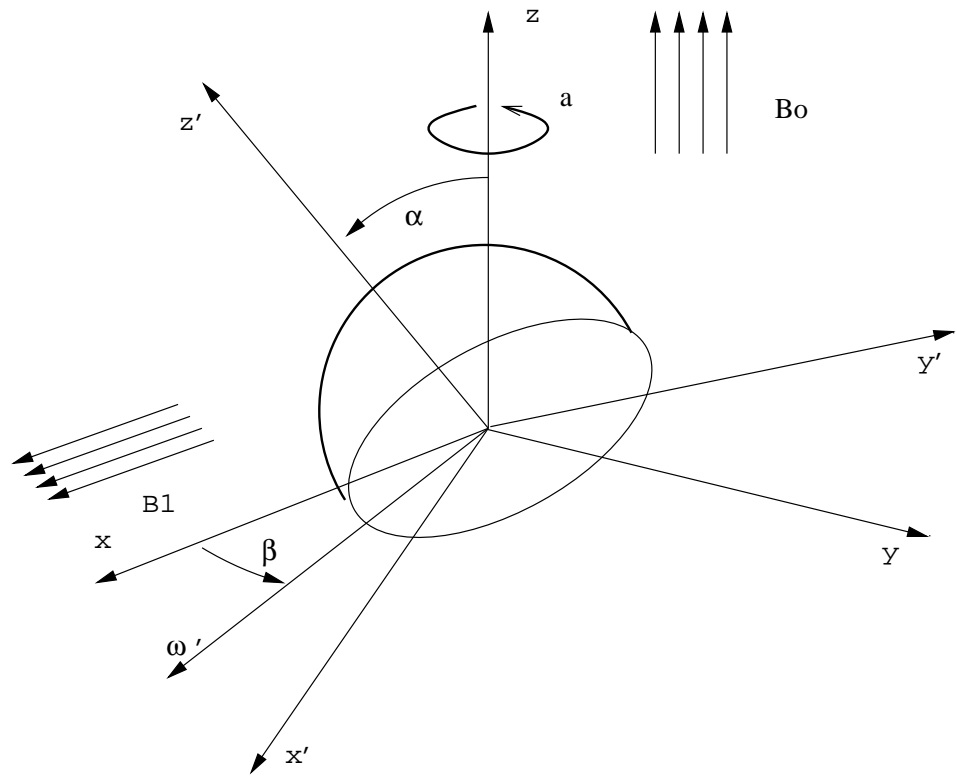
Fig.1. A generally located hemisphere on some part of the rotating hole's horizon
       across which the magnetic flux of an asymptotically uniform magnetic field
       is to be evaluated. Two angles $(\alpha, ~\beta)$ completely specify the
       location of the hemisphere and the hole rotates around the $z$-axis.
       Asymptotically, the magnetic field is decomposed into two components ;
       $B_{0}$, along the $z$-axis, the hole's axis of rotation and $B_{1}$, along
       the $x$-axis which is perpendicular to the hole's rotation axis.

\end{document}